\documentclass[11pt]{article}
\usepackage{moriond2000,epsfig}

\def\be{\begin{equation}}
\def\ee{\end{equation}}
\def\bea{\begin{eqnarray}}
\def\eea{\end{eqnarray}}

\begin{document}
\vspace*{3cm}
\title{DETECTION OF COSMIC SHEAR WITH THE WILLIAM HERSCHEL TELESCOPE}

\author{D. J. BACON, A. R. REFREGIER \& R. S. ELLIS}

\address{Institute of Astronomy, Madingley Road, Cambridge, CB3 0HA, England}

\maketitle\abstracts{Gravitational lensing by large-scale structure
induces weak coherent alignments in the shapes of background
galaxies. Here we present evidence for the detection of this `cosmic
shear' at the 3.4$\sigma$ significance level with the William Herschel
Telescope.  Analysis and removal of notable systematic effects, such
as shear induced by telescope optics and smearing by tracking and
seeing, are conducted in order to recover the physical weak shear
signal. Positive results for shear recovery on realistic simulated
data are presented, enhancing confidence in the measurement
method. The detection of cosmic shear is statistically characterised,
and its cosmological significance is discussed.}

\section{Introduction}

Understanding the large-scale distribution of matter in the universe
continues to be a major issue in modern cosmology. Weak gravitational
lensing promises to be a particularly effective method for determining
properties of large-scale structure, since it provides direct
information concerning the total mass distribution, independently of
its state and nature.

The images of distant field galaxies obtained at a telescope are
slightly coherently distorted, due to weak lensing by large-scale
structure. With extensive measurements of this shear on various
scales, one would obtain a direct measure of the power spectrum of
density fluctuations along the line of sight.

However, the first stage in such a programme is the detection of the
cosmic shear signal; this itself is a challenge, because the rms shear
amplitude is small - a few percent on arcminute scales. Recently four
papers describing the detection of this effect have been released
(Wittman et al 2000, van Waerbeke et al 2000, Bacon et al 2000a, Kaiser
et al 2000), presenting mutually consistent results with careful
analysis of systematic effects. 

Here we overview the detection of cosmic shear obtained with the 4.2m
William Herschel Telescope, fully discussed in Bacon, Refregier \&
Ellis (2000a). The current paper describes the survey strategy, and
discusses how the data are analysed to overcome convincingly the
contribution of systematic effects to the shear. Simulations used to
check our methodology are explained, and the cosmological implications
of our results are discussed.

\section{Observations}

The goal of our survey is to obtain a homogeneous sample of deep
fields, chosen to be on random lines of sight separated by
$>5^\circ$ in order to sample independent structures. The
galactic latitude of the fields was tuned to afford $\simeq200$ stars
within the field of view, necessary to correct for anisotropic PSF
systematics.

We carried out deep $R$-band imaging on 14 such fields with the Prime
Focus camera on the WHT. This has an 8'$\times$16' field of view and
pixel size 0.237''. The fields were exposed for one hour in $R$; median
seeing was 0.81'', having excluded exposures with seeing $>$ 1.2''. We
reach a magnitude depth of $R_{\rm median} = 25.2$, with $R_{\rm
median}$ of 23.4 for our selected sample, $z_{\rm median}\simeq 0.8$, and a
number density $N = 14.3$ arcmin$^{-2}$. 

\section{Analysis of Systematic Effects}

The aim of our analysis is to remove carefully systematic effects from
the galaxies' measured ellipticities, leading to unbiased measures of
the small ($\simeq$1\%) mean shear components for each field. We wish
to measure the mean shear in 8'$\times$8' cells, for increased shear
signal and to allow cross-correlation tests.

We used the Kaiser, Squires \& Broadhurst (1995) method (KSB)
implemented by Kaiser's {\tt imcat} software for object detection and
shape measurement. A detection signal-to-noise $\nu>15$ limit was imposed on
our usable galaxy catalogue to remove correction systematics found at low
signal-to-noise level.

\begin{figure}
\begin{minipage}{8.0cm}
\center
\psfig{figure=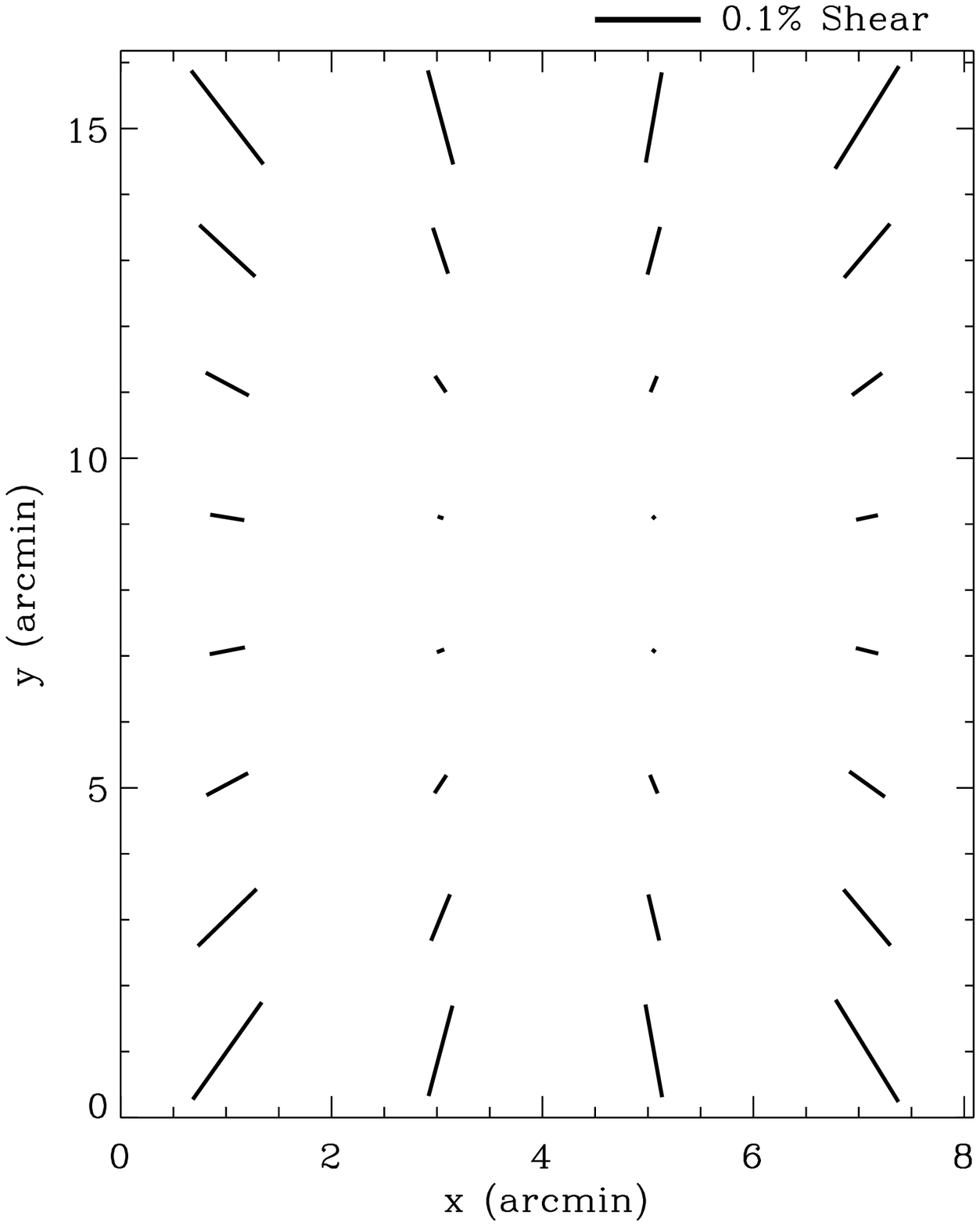,height=60mm}
\end{minipage}
\begin{minipage}{9.0cm}
\psfig{figure=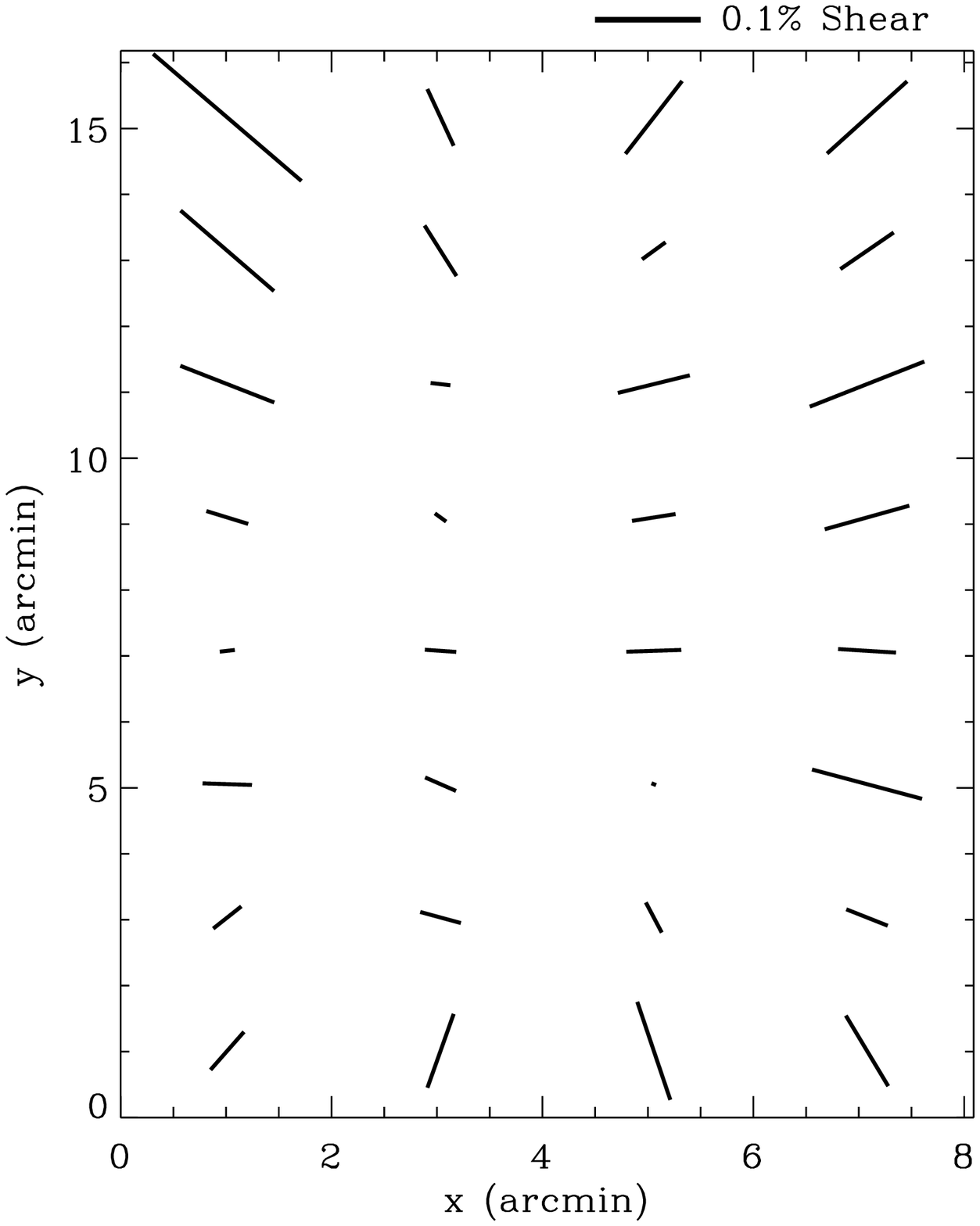,height=60mm}
\end{minipage}
\caption{Expected (left) and measured (right) instrumental shear
pattern for the WHT Prime Focus. The expected pattern was derived from
the distortion model given in the WHT Prime Focus manual (Carter \&
Bridges 1995). The observed pattern was measured using 3 astrometric
frames in one of our fields.
\label{fig:telshear}}
\end{figure}

The shear induced by telescope optics must be dealt with; by
calculating the telescope distortion using objects' relative positions
in several dithers, we show that this is negligible (shear due to
telescope $< 0.003$ everywhere; see figure \ref{fig:telshear}).

\begin{figure}
\begin{minipage}{8.0cm}
\center
\psfig{figure=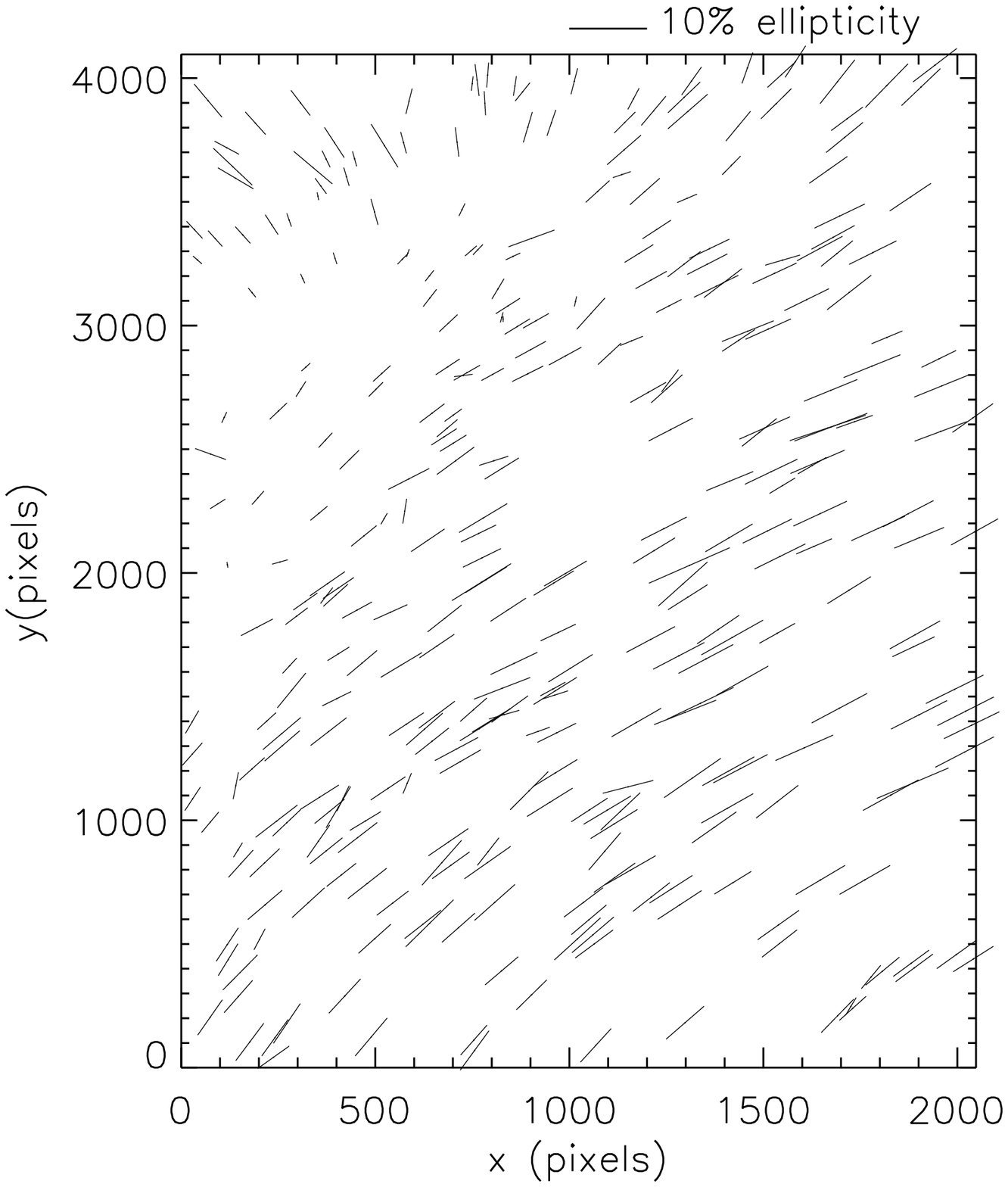,height=60mm} 
\end{minipage}
\begin{minipage}{9.0cm}
\psfig{figure=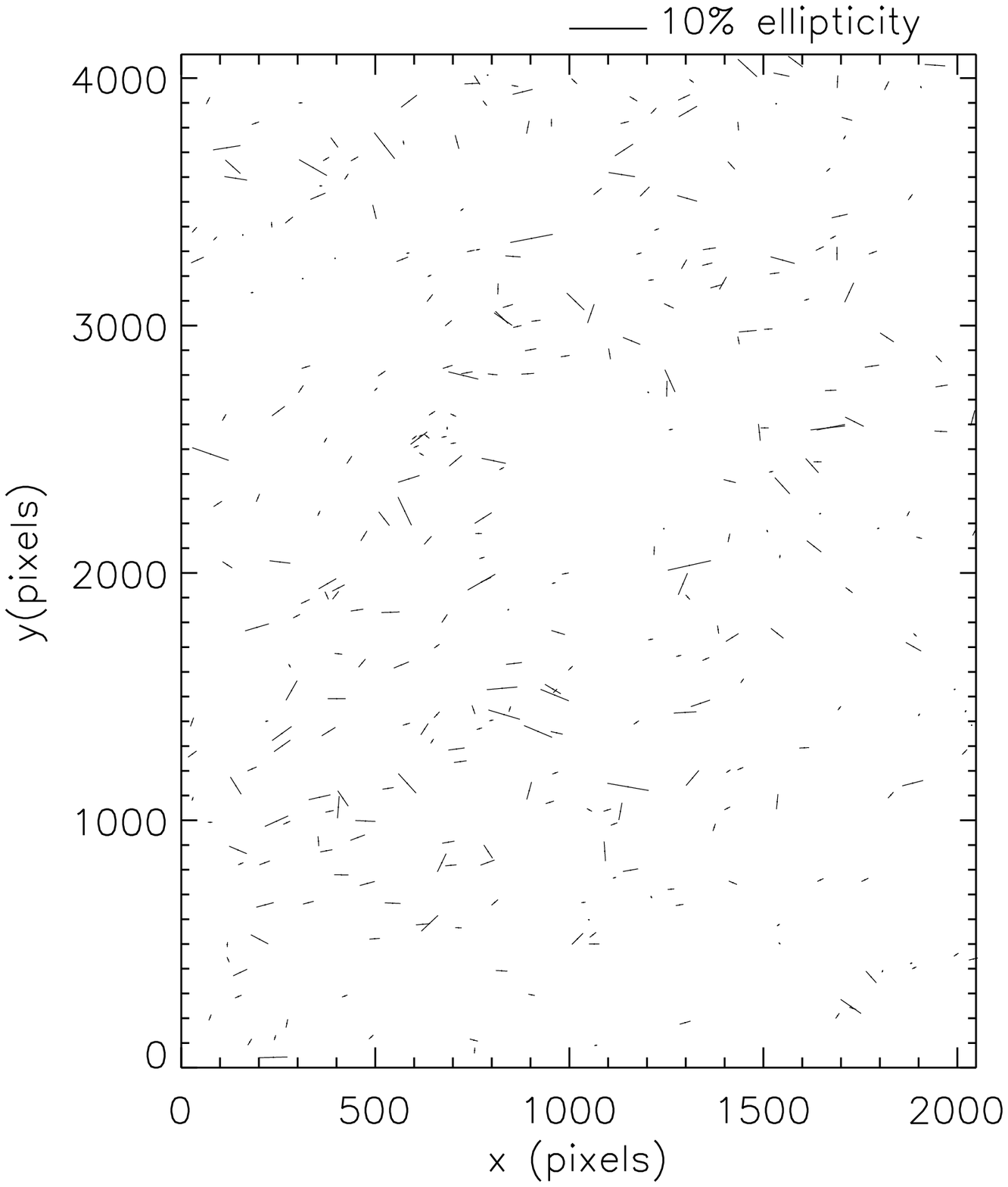,height=60mm}
\end{minipage}
\caption{Left: Stellar ellipticity distribution before correction for 
observed field (CIRSI2). Mean value observed $\bar{e}^{*}
\simeq 0.07$. Right: Residual stellar ellipticity after
correction. The residual mean ellipticity is $\bar{e}^{res} \simeq 2.6
\times 10^{-3}$.
\label{fig:smear}}
\end{figure}

Next, the PSF anisotropy from e.g. tracking errors must be removed
(see figure \ref{fig:smear}). Before correction this effect induces an
rms stellar ellipticity $e=0.07$, but after subtracting a fitted
2-dimensional cubic to stellar ellipticities, the residual stellar rms
is a mere $e = 1.4\times10^{-3}$. We correct the galaxy ellipticities
following Luppino \& Kaiser (1997), using the stellar fit model and
responsivities to smear measured for the galaxies.

Finally, galaxies are corrected for isotropic smear, i.e. the fact
that smaller galaxies' shapes have been more affected by
seeing-induced circularisation. Again we follow Luppino \& Kaiser's
(1997) method; this results in estimates for the mean shear components
in each 8'$\times$8' cell.

\section{Shear Measurements on Simulated Data}

In order to check our correction of systematics, and to calibrate
KSB-measured shear to real shear, we constructed simulated WHT fields
on which to carry out the above shear analysis (Bacon et al 2000b);
one can apply a chosen shear to a field, and test its recovery by our
algorithm. By creating a joint probability model of the magnitude -
number density - ellipticity - radius distribution of galaxies in
Ebbels' (1998) HST Groth Strip survey catalogue, we were able to draw
out statistically similar simulated catalogues for shearing and
analysis.

The catalogues were visualised with IRAF artdata; telescope-specific
pixelisation, throughput, anisotropic PSF, poisson and readout noise
were added. A null set of 20 simulated fields without shear were
created, together with a further 30 fields with an rms shear of
1.5\%. The KSB analysis described above was carried out on each field.

\section{Results}

\begin{figure}
\psfig{figure=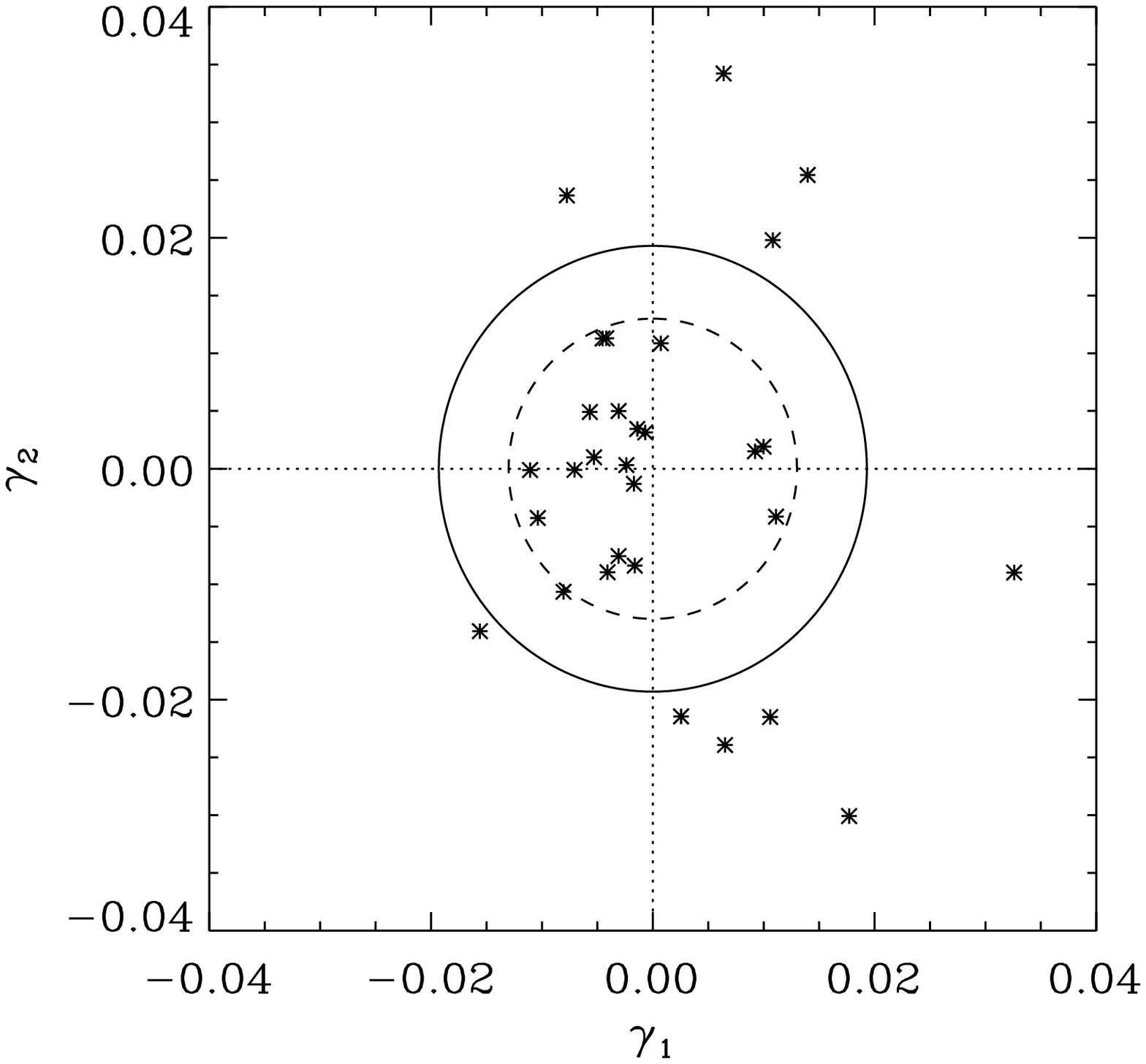,width=52mm}
\psfig{figure=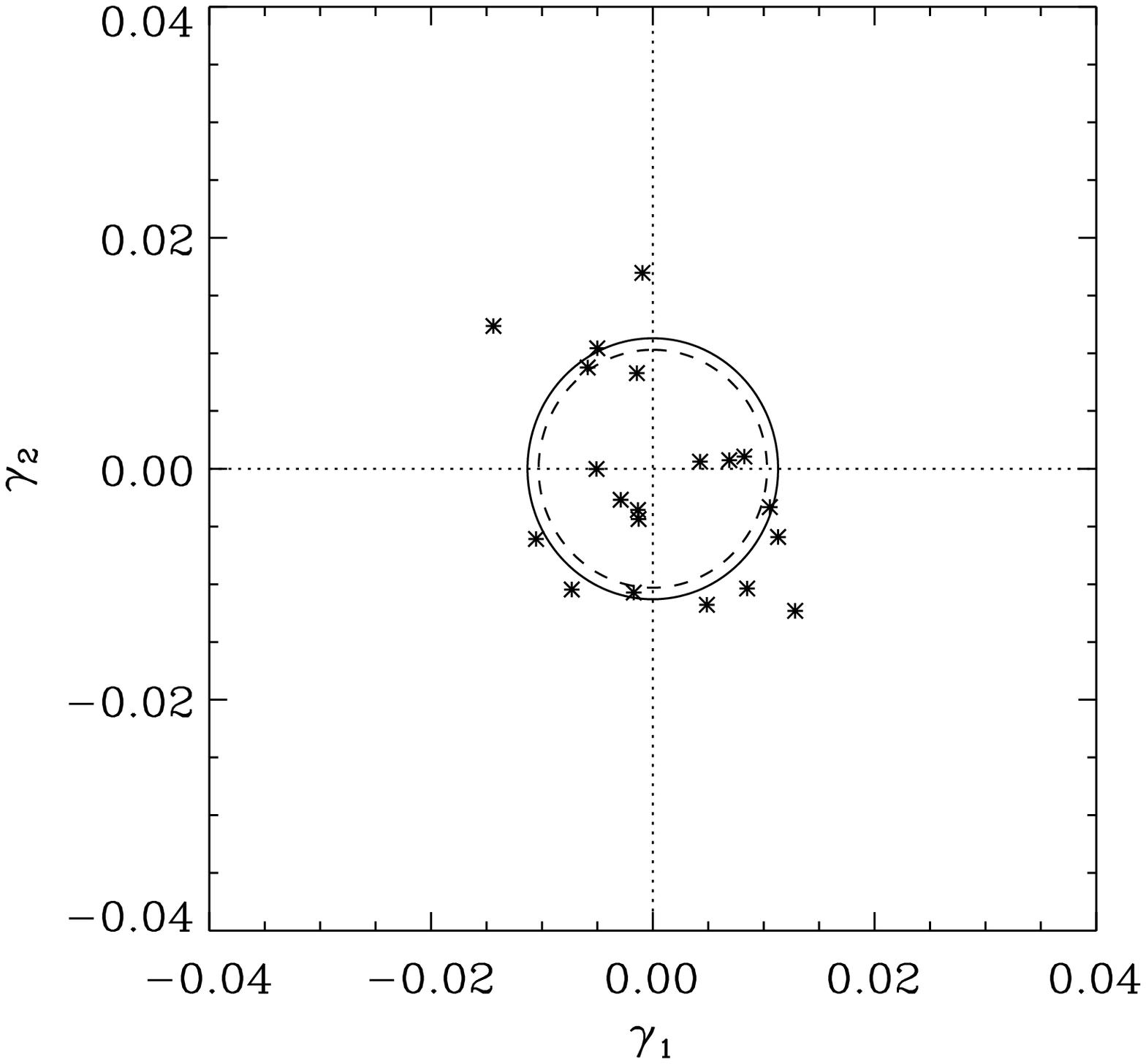,width=52mm} 
\psfig{figure=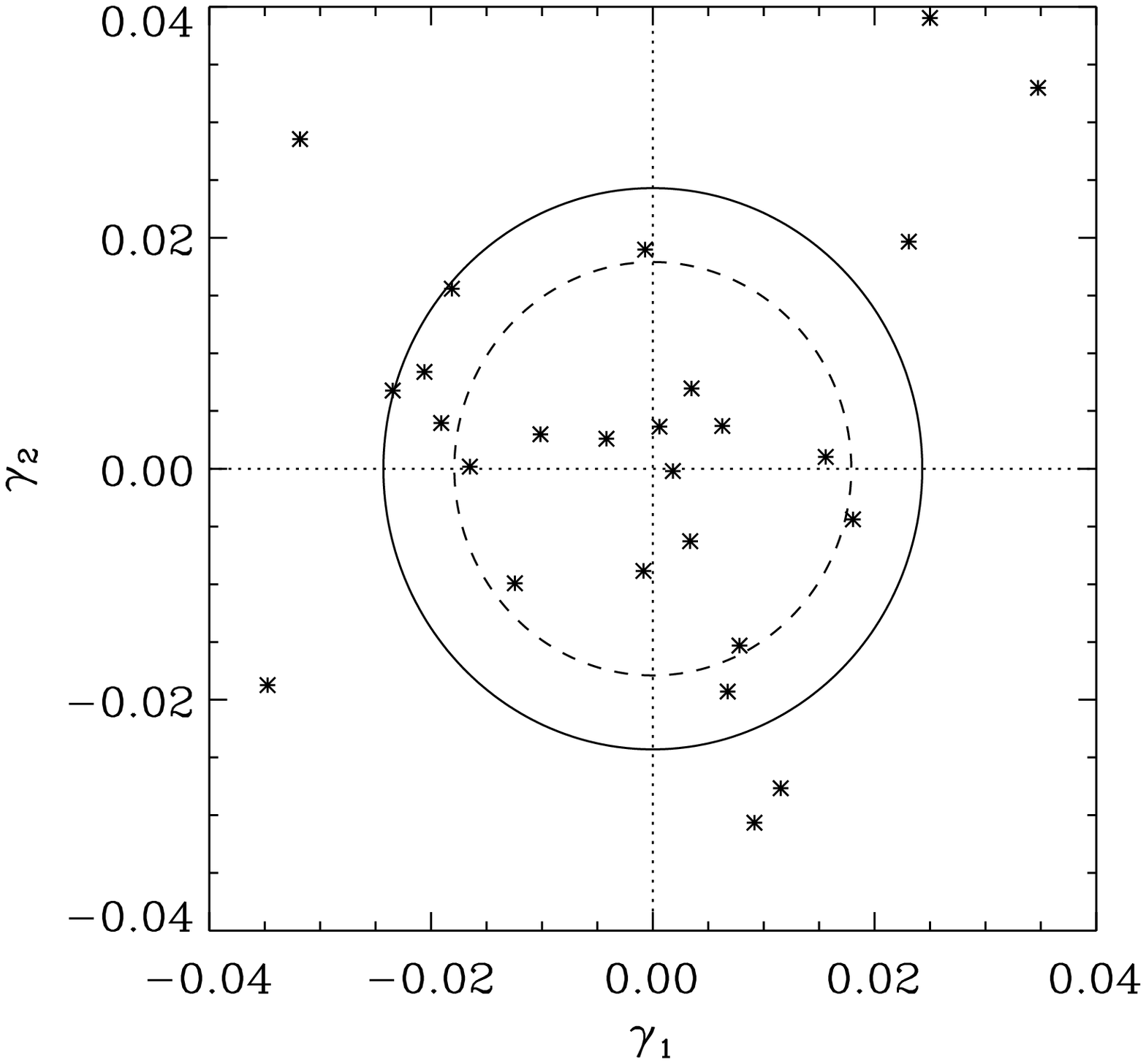,width=52mm}
\caption{Mean $\gamma_1$ and $\gamma_2$ for: (left) 30 simulated cells with
rms 1.5\% shear; (centre) 20 simulated null cells; (right) 26 observed
cells. The dashed circle shows the noise rms, the solid circle shows
the total rms. In the null case, the total rms is consistent with
noise alone; in the other cases the excess shear variance is
significant.
\label{fig:results}}
\end{figure}

We shall now compare the detection results for the simulated and
real data. The left-hand panel of figure
\ref{fig:results} shows the mean shear components found for our 1.5\%
rms shear simulations (30 cells). The inner circle represents the variance
that would be expected from noise alone; one can see that there is an
excess variance $\sigma_{\rm lens}^2$, which turns out to be
significant; $\sigma_{\rm lens}^2=(0.013)^2 \pm (0.006)^2$, to be
compared with our input $\sigma_{\rm lens}^2=(0.015)^2$. The fact that
the method has detected the simulated signal with the correct
amplitude is encouraging.

The middle panel of figure \ref{fig:results} shows the mean shear
components for 20 simulations with no shear added. Note that here the
variance is accounted for by noise alone; as expected, there is no
excess variance.

The shear results for our observed fields are shown in the right-hand
panel of figure \ref{fig:results}. Again we see an excess variance;
with a thorough statistical analysis we find this to be significant,
with $\sigma_{\rm lens}^2=(0.016)^2 \pm (0.008)^2 \pm (0.005)^2$, the
errors being due to noise and uncertainty on any remaining systematics
respectively. This corresponds to a 3.4$\sigma$ detection of cosmic
shear.

By comparing our $\sigma_{\rm lens}^2=(0.016)^2 \pm (0.012)^2$ (error
now includes cosmic variance) with that expected for popular
cosmological models, we find that COBE-normalised SCDM is ruled out at
3$\sigma$ level, whereas cluster-normalised $\tau$-, $\Lambda$-,
and OCDM are highly consistent with the data.

Our results also afford us a measure of $\sigma_8$ for a given
cosmological model. For instance, for $\Lambda$CDM with $\Omega_m$ = 0.3,
we obtain $\sigma_8=1.47 \pm 0.51$, consistent with cluster
abundance determinations, $\sigma_8=1.13 \pm 0.19$ (Viana \& Liddle 1996).

With increased numbers of fields in the future, cosmic shear variance
measurements will afford very precise estimates of $\sigma_8$, while
skewness measurements of the distortion field will provide an
independent estimate of $\Omega_m$.

\section{Conclusion}
Evidence for the detection of shear arising from large-scale structure
has been presented based on an analysis of 14 fields obtained at the
William Herschel Telescope. Particular attention has been paid to
questions of systematic correction and testing of measurement method
by simulations. Prospects are now bright for measuring with greater
accuracy the amplitude of the cosmic shear signal. The key
uncertainties to overcome are noise and cosmic variance, i.e. more
independent lines of sight will lead to a better estimate of the shear
amplitude. Future cosmic shear surveys will consequently provide
powerful constraints on cosmological parameters.

\section*{Acknowledgments}
Particular thanks go to Nick Kaiser for providing the Imcat software,
and to Douglas Clowe for valuable discussions as to its use. We also
thank Roger Blandford, Chris Benn, Andrew Firth, Mike Irwin, Konrad
Kuijken, Peter Schneider, Andrew Liddle, Yannick Mellier, Roberto
Maoli and Jason Rhodes for useful discussions. Max Pettini kindly
provided us with one of the WHT fields. This work was performed within
the European TMR lensing network.

\end{document}